\documentclass[conference]{IEEEtran}
\usepackage{epsfig,wrapfig,boxedminipage,amssymb,amsmath,amscd}
\usepackage{capt-of}\usepackage{times} 
\usepackage{epstopdf}
\usepackage[T1]{fontenc}
\usepackage[usenames]{color}
\usepackage[official,right]{eurosym}

\def\A{{\cal A}}

\def\comp{\raise 1pt \hbox{$\scriptstyle\circ$}}

\def\argmax{\mathop{\rm argmax}\limits}

\def\upto{{\raise 1pt \hbox{$\scriptstyle \,\nearrow\,$}}}
\def\downto{{\raise 1pt \hbox{$\scriptstyle \,\searrow\,$}}}

\IEEEoverridecommandlockouts    

\begin{document}
\newcommand{\boldx}{\mbox{$\mathbf{x}$}}
\newcommand{\boldy}{\mbox{$\mathbf{y}$}}
\newcommand{\boldf}{\mbox{$\mathbf{f}$}}
\newcommand{\boldz}{\mbox{$\mathbf{z}$}}
\newcommand{\boldF}{\mbox{$\mathbf{F}$}}
\newcommand{\boldG}{\mbox{$\mathbf{G}$}}
\newcommand{\boldg}{\mbox{$\mathbf{g}$}}
\newcommand{\boldh}{\mbox{$\mathbf{h}$}}
\newcommand{\boldH}{\mbox{$\mathbf{H}$}}
\newcommand{\boldzero}{\mbox{$\mathbf{0}$}}
\newcommand{\Rbb}{\mbox{$\mathbb R$}}

\newcommand{\boldq}{\mbox{$\mathbf{q}$}}
\newcommand{\boldtau}{\mbox{\boldmath$\tau$}}

\newtheorem{defn}{Definition}

\title{Multi-objective Stackelberg Game Between a Regulating Authority and a Mining Company: \\A Case Study in Environmental Economics}
\author{Ankur Sinha, Pekka Malo, Anton Frantsev and Kalyanmoy Deb$^*$\thanks{$^*$Also Department of Mechanical Engineering, Indian Institute of Technology Kanpur, PIN 208016, India (deb@iitk.ac.in).}\\
Department of Information and Service Economy\\
Aalto University School of Business\\ 
PO Box 21220, 00076 Aalto, Helsinki, Finland\\
{\{Firstname.Lastname\}@aalto.fi}
}
\maketitle

\begin{abstract}
Bilevel programming problems are often found in practice. In this paper, we handle one such bilevel application problem from the domain of environmental economics. The problem is a Stakelberg game with multiple objectives at the upper level, and a single objective at the lower level. The leader in this case is the regulating authority, and it tries to maximize its total tax revenue over multiple periods while trying to minimize the environmental damages caused by a mining company. The follower is the mining company whose sole objective is to maximize its total profit over multiple periods under the limitations set by the leader. The solution to the model contains the optimal taxation and extraction decisions to be made by the players in each of the time periods. We construct a simplistic model for the Stackelberg game and provide an analytical solution to the problem. Thereafter, the model is extended to incorporate realism and is solved using a bilevel evolutionary algorithm capable of handling multiple objectives. 

\end{abstract}


\section{Introduction}
\label{sec:intro}
We present and solve a multi-objective Stackelberg competition model found in the field of environmental economics, where the aim of a regulating authority is to earn revenues through taxes and regulate the environmental damages caused by a gold mining company. The regulating agency in such a problem is usually the government which acts as a leader. The mining firm is the follower, which reacts rationally to the decisions of the leader in order to maximize its own profit. In this problem, the leader has two objectives while the follower has one. In this strategic game, the leader solves the problem to find his optimal strategy, assuming that he possesses all necessary information about the follower. This gives rise to two levels of optimization tasks, with one optimization problem nested within the other. Such problems are also commonly referred to as bilevel optimization problems \cite{bilevel-book} in the literature.

The problem in this paper involves finding optimal actions for the government over multiple time periods for which the mine operates, with one objective being overall tax revenue maximization, and the other being preservation of the environment by way of pollution minimization. The actions of the leader consist of choosing the optimal tax structure. The follower chooses an optimal technology to match the tax structure as well as makes optimal extraction decisions in order to maximize the expected profits. Such a problem frequently arises for authorities making environmental regulatory decisions. Due to the difficulties involved in handling such environmental problems, the authorities usually choose a satisfying solution, instead of solving the problem to optimality.

There have been many studies concerning bilevel optimization problems \cite{colson,vicente-review,dempe-dutta}, and their practical applications are well documented in the literature \cite{bilevel-book}. 
Such problems differ from common optimization problems because they contain a nested optimization task within the constraints of the outer problem. The main optimization task is usually termed as the upper level problem, and the nested optimization task is referred to as the lower level problem. Due to the tiered structure of the overall problem, a solution to the upper level problem may be feasible only if it is also an optimal solution to the lower level problem. This requirement makes finding a solution to bilevel optimization problems particularly challenging. 

Bilevel problems are often solved using approximate solution methodologies, which may not necessarily lead to an optimal solution \cite{bianco-kkt,bilevel-book,stackelberg-design} for a complex case. Some common techniques employed by researchers and practitioners in handling such problems include the Karush-Kuhn-Tucker approach \cite{bilevel-KKT1,bianco-kkt}, the use of penalty functions \cite{aiyoshi81}, and Branch-and-bound techniques \cite{bard82}. Further useful insight and discussion on the subject of bilevel programming may be found in \cite{colson} and \cite{vicente-review}. Another approach for handling complicated bilevel optimization problems are evolutionary algorithms, which have proven to be potent tools for handling such tasks in the recent past \cite{yin-bilevel,GA_Wang}.

The presence of multiple objectives appends further challenges to a bilevel programming problem. The difficulties involved in handling such problems have deterred researchers from working on effective methodologies for solving multi-objective bilevel problems. Practical applications are also rare because practitioners often tend to pose such problems with a single objective, though the inherent nature of the problem might involve multiple objectives. Despite these challenges recently there has been interest in these problems \cite{halter-sanaz,shi-xia,my-ecj10,my-ifac09,ruuska12,zhang12}.

The paper is structured in the following manner. In Section \ref{sec:desc} we provide a description of a multi-objective bilevel problem. Section \ref{sec:case} outlines the case study of the paper. Section \ref{sec:model} develops the model used in this study by first providing a closed-form solution to a simple formulation of the problem and then by presenting and solving an extended dynamic model with more realism. Section \ref{sec:algorithm} describes the solution procedure employed to solve the extended model. Section \ref{sec:results} discusses the results obtained through the solution process. Section \ref{sec:conclusions} summarizes and concludes the paper.

\section{Description of a Multi-objective Bilevel Problem}
\label{sec:desc}
A general multi-objective bilevel optimization problem has two levels of multi-objective optimization tasks. However, the problem considered in this paper has two objectives at the upper level, and only a single objective at the lower level. Such a multi-objective bilevel optimization problem can be described as follows:
\begin{equation}
\begin{array}{rl}
\mbox{Max}_{(\boldx_u,\boldx_l)} & \boldF(\boldx) =
\left(F_1(\boldx),F_2(\boldx)\right), \\
\mbox{subject to} & 
\boldx_l \in \argmax_{(\boldx_l)}\left\lbrace 
 f(\boldx) \big| \boldg(\boldx) \ge \boldzero, \boldh(\boldx) = \boldzero
	 \right\rbrace, \\
& \boldG(\boldx) \ge \boldzero, \boldH(\boldx) = \boldzero,\\
& x_i^{(L)} \le x_i \le x_i^{(U)}, \quad i=1,\ldots,n. 
\end{array} 
\label{eq:bilevel_multi_obj}
\end{equation}

In the above formulation, $F_1(\boldx), F_2(\boldx)$ are
upper level objective functions, and $f(\boldx)$ is the lower level objective function. 
The functions $\boldg(\boldx)$ and $\boldh(\boldx)$ determine the feasible space
for the lower level problem. The decision vector is $\boldx$ which comprises of two
smaller vectors $\boldx_u$ and $\boldx_l$, such that $\boldx = (\boldx_u, \boldx_l)$.
The lower level
optimization problem is optimized only with respect to the variables
$\boldx_{l}$, while the variables $\boldx_{u}$ act as fixed parameters for the
problem. Therefore, the solution set of the lower level problem can be
represented as a function of $\boldx_u$, or as $\boldx_l^{\ast}(\boldx_u)$. This means that the upper level variables ($\boldx_u$) act
as a parameter to the lower level problem, and hence the lower level optimal solutions ($\boldx_l^{\ast}$) are a
function of the upper level vector $\boldx_u$.
The functions $\boldG(\boldx)$ and
$\boldH(\boldx)$ along with the optimality to the lower level
problem determine the feasible space for the upper level optimization
problem. 

\section{Mining vs Environmental Damage: A Case Study}
\label{sec:case}
Kuusamo region lies in the northern part of Finland. It is well known for its natural beauty and is a popular tourist resort. Recently, there has been a lot of interest in this region, as it is considered to be a ``highly prospective Palaeoproterozoic Kuusamo Schist Belt'' \cite{DragonMining}, 
which contains large amounts of gold deposits. One of the companies, Dragon Mining \cite{DragonMining}, which is an Australia-based company primarily operating in the Nordic region has been performing drill tests to evaluate the mining prospects. The average gold content in the ore is expected to be around 4.9 grams per ton \cite{HS1},  which is worth millions of Euros considering the overall deposits present in the area. Though the mining project would lead to a large amount of gold resources and also generate a number of jobs, it is being opposed for the fear of the harm which it might cause to the environment. There are three main reasons for the opposition against the gold mining operations in Kuusamo. Firstly, the river Kitkajoki is located in Kuusamo, and the environmentalists fear that the run-off water generated from the gold mining operations might pollute the river water. Secondly, the ore contains uranium, which if mined, would blemish the reputation of the tourist resort. Thirdly, the visible open-pit mines located next to the Ruka slopes, will be a big turn-off for skiing and hiking enthusiasts. 

Under such a situation, it is the onus of the government to make a decision, whether to allow mining and to what extent. The government here has primarily two objectives: the first objective is to maximize the revenues generated by the mining project, which may include the additional jobs, taxes, etc; and the second objective is to minimize the harm caused to the environment as a result of mining. Obviously, there is a trade-off between the two objectives, and the government as a decision maker needs to choose one of the preferred trade-off solutions. The government is aware that the mining company has a sole objective of maximizing its profit under the constraints set by the government. In this scenario, the government would like to have a tax structure such that it is able to maximize its own revenues in addition to being able to restrain the mining company from causing extensive damage to the environment. There is a hierarchy in the problem, which arises from the manner in which the two entities operate. The government has higher control of the situation and decides the terms and conditions for the mining company to operate in. Therefore, in this framework, we observe that the government acts as a leader, and the mining company acts as a follower.

As a leader, it is possible for the government to optimally regulate the problem in its favour, provided that it has complete knowledge of the follower's strategies. If the government decides to tax the mining company based on each unit of gold it produces, then for any given tax structure, the mining company will solve its own optimization problem to maximize its profit. However, if the government already takes the mining company's optimization task into account, then it would be possible for them to generate their own optimal strategies. The overall problem appears as a bilevel optimization task, where for each tax structure, the government observes how the company acts and then chooses that particular tax structure which suits it the most, taking the actions of the mining company into account. It should be noted that in spite of different objectives appearing in the problem, it is not possible to handle such a problem as a simple multi-objective optimization task. The reason for this is that the leader cannot evaluate any of its own strategies without knowing the strategy of the follower, which it obtains only by solving a nested optimization problem.

The model considered in this paper does not attempt to resolve the decision of establishing
the mine or to argue for the benefits or detriments of any decision taken by the government. What it does attempt to produce is an optimal set of decisions for the government from which it might want to choose the most preferred one. 

\section{Problem Formulation}
\label{sec:model}

In this section, we present the models used in our analysis. We begin by showing an analytical solution to a simple multi-objective bilevel optimization problem. This model demonstrates the procedure for solving problems of this type as long as the functions are continuous, differentiable and convex. In the following subsection, we present an extended version of this model. This larger model incorporates a dynamic aspect and several other improvements, like alternative technologies and non-linear cost functions, which make the model more realistic and therefore more representative of the actual problem.

In both parts, the described situation is viewed from the perspective of the regulating entity at the upper level, which we will refer to as the ``government'', which acts much like the leader in a classic Stackelberg competition. The follower in this setting is the mining firm, which is commonly referred to throughout the paper as simply the ``mine''. Since the problem has multiple objectives at the upper level, the solution of the bilevel problem leads to a Pareto-optimal frontier. After having obtained the optimal trade-off solutions, the government is faced with a multi-criteria decision making problem, whereby it must effectively balance the optimal revenue it receives from taxation with the reduction in overall welfare caused by the pollution. The leader's choice for higher taxes and pollution reduction conflicts with the follower's choice of profit maximization, leading to a Stackelberg competition with the leader having the first mover's advantage.


\subsection{Basic Analytical Model}
\label{sec:basicModel}

We start with the simple problem as viewed by the government. The government tries to maximize overall welfare by imposing a tax on the mine and collecting the largest possible tax revenue while at the same time trying to minimize the amount of pollution produced by the mine. This problem can be described as follows.
\begin{align}
\max_{\tau,q} \quad &\boldF(q, \tau) = (R, - D) \label{eq:object} \\
\mbox{s.t.} \quad & q \in \argmax_{q}
			   \left\{ \begin{aligned}
			    \pi(q)& = p(q)q - c(q) - R\\
			    \pi(q)& \ge 0	\\
			  \end{aligned} \right\} \label{eq:constr1}\\
& q \ge 0, \tau \ge 0.
\end{align}

In (\ref{eq:object}), the first objective deals with the tax revenue, where $R = \tau q$; $\tau$ is the per unit tax imposed on the mine, and $q$ is the amount of metal extracted from the ore by the latter. The second objective denotes the environmental damage caused by the mine that the government ultimately wants to minimize. $D = kq$, where $k$ is the pollution coefficient signifying the negative impact of extraction on the environment. The damages are thus linear and scale proportionately with the amount of gold extracted from the earth since a larger base of operation implies larger environmental damage.

\begin{figure}
		\includegraphics[width=\linewidth]{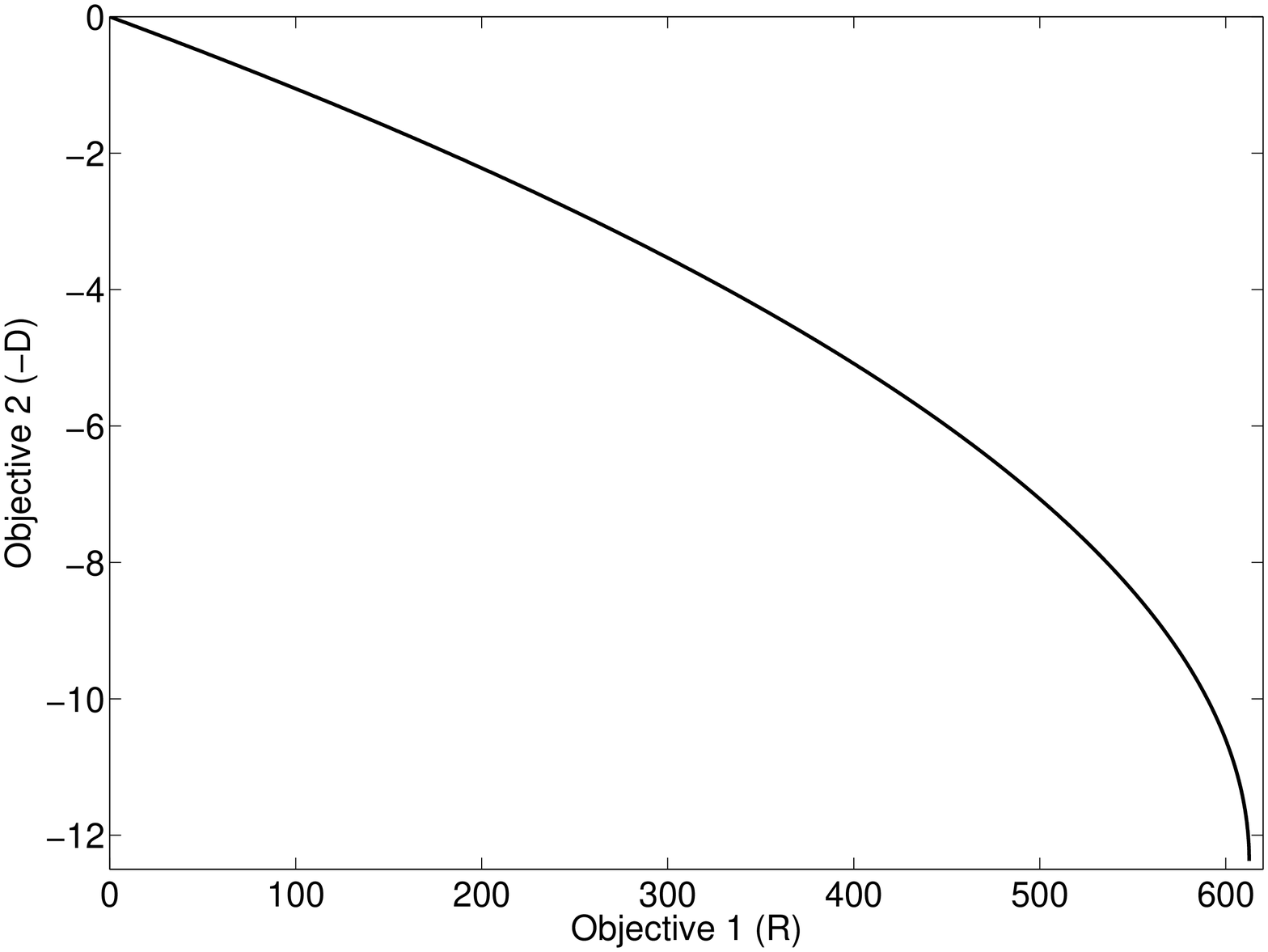}
		\caption[Pareto-optimal Front]{Pareto-optimal frontier for the government showing the trade-off between tax revenues and environmental pollution.}
		\label{fig:analytical}
\end{figure}

\begin{figure}
		\includegraphics[width=\linewidth]{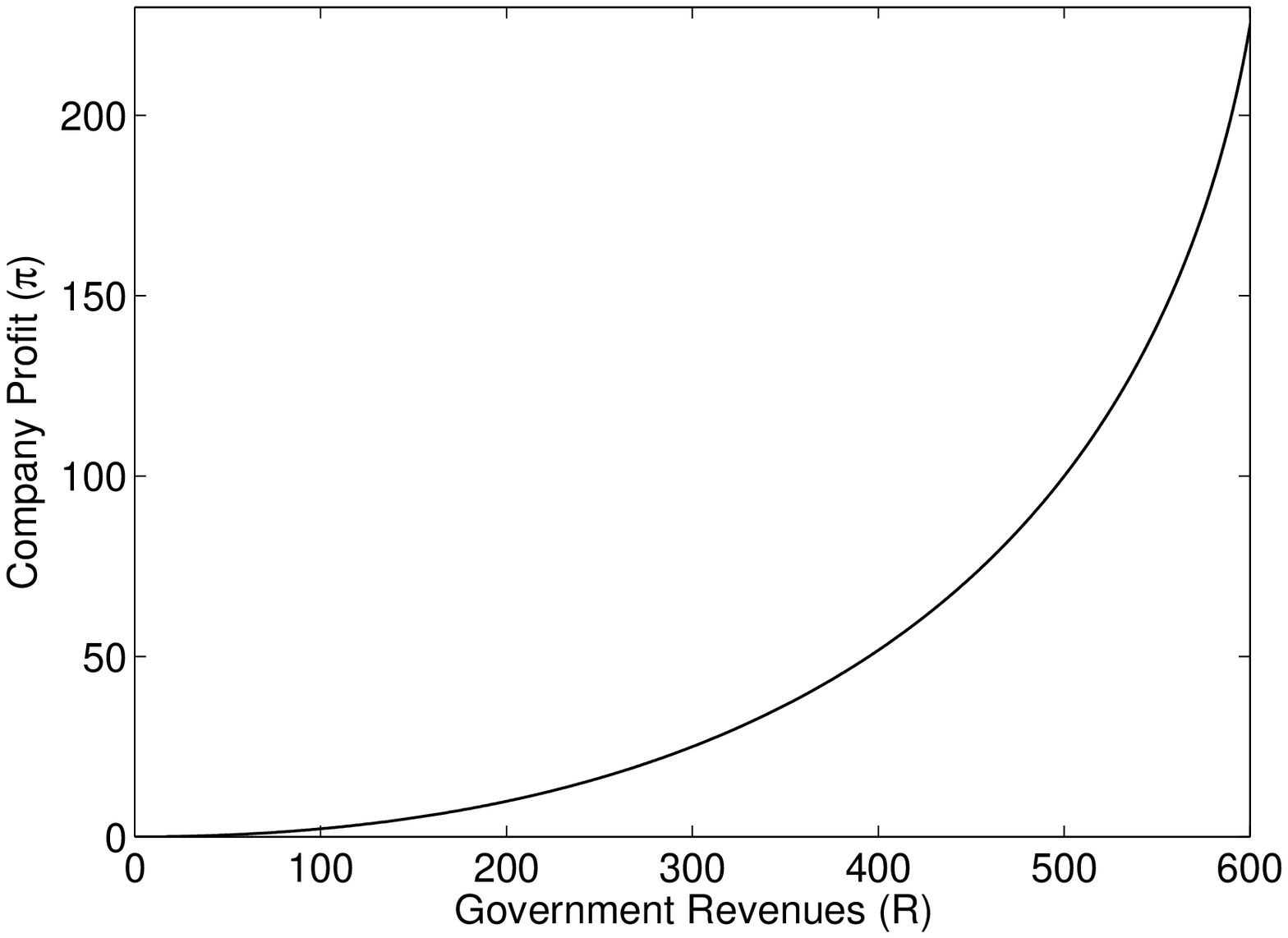}
		\caption[Sharing]{Revenues earned by the government and the profit of the operating company for the Pareto-optimal instances.}
		\label{fig:analytical2}
\end{figure}

Equation (\ref{eq:constr1}) gives the profit of the mine, where $p(q)q$ (price function times amount of metal extracted) is the revenue function, and $c(q)$ is the extraction cost function followed by the additional tax levied on the mine. The mine is most likely to be a price taker when it comes to the price of gold and must base its mining decisions on the possible price paid by their customers. It would therefore be plausible to replace the price function for gold in the above equation by a constant. However, given the assumption that the mine can extract a large amount of ore, and subsequently gold, at one time, it would be possible for it to affect the price of gold slightly. Therefore, we assume the price function to be linear with a small slope. Extraction cost is considered to be quadratic since extracting ore that lies deeper underground tends to get increasingly expensive. Thus, we have the following model:
\begin{flalign}
\max_{\tau,q} & \quad \boldF(q, \tau) = (\tau q, - kq) \label{eq:object2}\\
\mbox{s.t.} & \quad \notag\\
& q \in \argmax_{q} 
			   \left\{ \begin{aligned}
			    \pi(q)=&(\alpha - \beta q)q - \\ &(\delta q^{2} + \gamma q + \phi) - \tau q\\
			    \pi(q) \ge& \hspace{1mm} 0	\\
			  \end{aligned} \right\}\\
&q \ge 0, \tau \ge 0,
\end{flalign}
where $\alpha, \beta, \delta, \gamma, \phi$ are constants, and $\phi$ represents the fixed costs of setting up operations. The parameters are chosen as $\alpha=100, \beta=1, \delta=1, \gamma=1$ and $\phi=0$. 
It is in the interest of the leader that the follower always extracts ore from the mine. If the follower does not make any extraction, then the revenues earned by the leader will also be zero. When the parameter $\phi$ is zero, the constraint at the lower level is always satisfied.
Therefore, we solve the follower's optimization problem in an unconstrained manner. By applying the standard First-Order Approach, we find the points at which the follower's profit is maximized.
\begin{equation}
	\frac{d \pi}{d q} = \alpha - 2\beta q - 2\delta q - \gamma - \tau = 0.
\end{equation}
Simply rearranging some terms in this equation yields the optimum expression for the extraction amount, $q$, in terms of the tax variable, $\tau$:
\begin{equation}
	q = q(\tau) = \frac{\alpha - \gamma - \tau}{2(\beta + \delta)}.
\end{equation}
We attempt to solve the multi-objective problem at the upper level using a weighted sum approach. Therefore, we frame the government's weighted single objective function as
\begin{equation}
F(q, \tau, w) = w\tau q - (1-w)kq,
\end{equation}
where $(w,1-w)$ represents the weights for each of the objectives.
Inserting the expression for $q$ into the government's optimization problem and rearranging, we get the following:
\begin{equation}
F(\tau, w) = \left(w \tau + w k - k\right) \frac{\alpha - \gamma - \tau}{2(\beta + \delta)}.
\end{equation}
Following the same procedure as we used for the lower level optimization problem, we set the first-order differential of the leader's problem to zero and solve for the optimal tax rate.
\begin{equation}
\frac{d F}{d \tau} = \frac{w(\alpha - \gamma - \tau)}{2(\beta + \delta)} 
			- \frac{w \tau + w k - k}{2(\beta + \delta)} = 0.
\end{equation}
Combining like terms, simplifying, and rearranging slightly yields the optimal choice of tax, $\tau^*$, in terms of the constant parameters of the model and the government's preference parameter, $w$.
\begin{align}
2w \tau &= w(\alpha - \gamma - k) + k ,\\
\tau^*(w) &= \frac{\alpha - \gamma - k}{2} + \frac{k}{2w}.
\end{align}
%
We can now replace the tax parameter in the mine's optimization problem by its respective representation above and solve for the optimal output of the mine.
\begin{equation}
q = \frac{\alpha - \gamma}{2(\beta + \delta)} 
	- \frac{\alpha - \gamma - k}{2 \cdot 2(\beta + \delta)} 
	- \frac{k}{2w \cdot 2(\beta + \delta)}.
\end{equation}
Again, combining like terms and rearranging slightly gives us the optimal extraction quantity of the mine, $q^*$, in terms of the constant parameters of the model and the government's preference weight, $w$.
\begin{align}
q^*(w) &= \frac{w(\alpha - \gamma)-(1-w)k}{4w(\beta + \delta)}\label{eq:production}.
\end{align}
%
Therefore, the government can influence the extraction rate of the mine based on its own preferences for tax revenue versus environmental conservation, if it has information about the mine's costs. By varying the government's preference weights, it is possible to generate the entire Pareto-optimal frontier for the multi-objective bilevel problem. From (\ref{eq:production}), after substituting the parameter values we find that $w < 0.01$ generates infeasible solutions ($q<0$). Therefore, the Pareto frontier is generated using weights $0.01 \le w \le 1$. We present the Pareto-optimal frontier for this simple model in Figure \ref{fig:analytical}. Figure \ref{fig:analytical2} represents the revenues of the government and the profit of the company. We observe that with increasing damage to the environment revenues of the government as well as profit of the company rises.

%
\subsection{Extended Multi-Objective Model}
\label{sec:extendedModel}

We next consider an extended version of this model. The main differences between this model and the simple model in the previous section are incorporation of multiple time-periods, variable price for different periods, inclusion of several different technology options for the mine to choose from, and incorporation of different types of costs. Solving the problem in multiple time-periods leads to a significant increase in the number of variables, choice of different technology options introduces discreteness into the problem, and incorporation of complex cost functions introduces non-differentiability. With these extensions, it is very difficult to handle the problem using an analytical approach or a classical optimization algorithm. Therefore, after providing a description of the extended model and the complexities, we discuss the solution procedure used to handle this model.

We keep most of the notation from the previous simple model the same with only a few exceptions. With the addition of more time periods into the model, a subscript $t = 1, \dotsc, T$ is added to the variables that potentially change from one period to the next. The mine's only endogenous variable in each period is the decision of the extraction amount. Thus, a strategy of the mine for the duration of the game is then given as $\boldq = \left(q_1, q_2, \dotsc, q_T\right)$. The government influences the mine by varying the tax rate. Therefore, a strategy for the government for the duration of the game is given by $\boldtau = \left(\tau_1, \tau_2, \dotsc, \tau_T\right)$. Most of the literature focuses around a tax rate that is kept constant by the government once set \cite{Conrad1981b,Conrad1981a,Helliwell1978}. However, in such a framework, a constant tax rate might not be an optimal strategy, and it would be appropriate for the government to vary the tax rate throughout the life of the mine. The technology alternatives are represented by a discrete variable $a$. Therefore, the pollution coefficient $k$ is a function of $a$; $k$ is larger for a technology which causes more damage to the environment, and smaller for the technology which causes less damage to the environment.
Thus, the multi-period multi-objective bilevel optimization problem can be presented as follows.

\begin{align}
\max_{\boldtau,\boldq,a} \quad &\boldF(\boldq, \boldtau,a) = \left(\sum_{t=1}^{T} \tau_{t} q_{t}, -\sum_{t=1}^{T} k(a) q_{t} \label{eq:object3}\right)\\
\mbox{s.t.} \quad & \boldq \in \argmax_{\boldq,a} \left\{ \Pi(\boldq,a) = \sum_{t=1}^{T} \pi_{t} (q_{t},a) \right\}, \label{eq:sumProfit}\\
& q_t \ge 0 \hspace{1mm} \forall \hspace{1mm} t \in \{1,\ldots,T\}, \sum_{t=1}^{T} q_{t} \leq S,
\end{align}
%
%
%
where $S$ is the overall available stock of the resource in the area. It is noteworthy that the stock constraint is irrelevant, as it is not profitable for the mine to extract the entire available stock because the extraction costs rise significantly. The costs increase with deep mining, and moreover the concentration of the gold in the ore becomes smaller when the majority of the stock is extracted, making any further extraction non-profitable. Therefore, we ignore the stock constraint hereafter.

The mine's profit function for each period, $\pi_{t}$, is defined as
\begin{align}
\pi_t (q_t,a) &= (\alpha_{t} - \beta_t q_{t})q_{t} - c_{i}^{er}(q_{t},a) \notag\\
&- c_{i}^{ep}(q_1,\ldots, q_t,a) - \tau_{t} q_{t} . \label{eq:profitMine3}
\end{align}
The first product term $(\alpha_{t} - \beta_t q_{t})$ in (\ref{eq:profitMine3}) represents the price function over time. The price function for gold is determined exogenously based on future market conditions and the amount of gold extracted by the company. This reflects our assumption that the price of gold may change from one period to another. Parameter $\alpha_t$ reflects the market conditions in a particular period, and parameter $\beta_t$ reflects the small impact on the price of gold caused by the extraction. The equation contains two cost terms, $c_{i}^{er}$ and $c_{i}^{ep}$. The first cost term represents the extraction rate cost, i.e. rate at which the company extracts the ore in a particular period. This term restricts the company from extracting the entire ore in a single period, as it is practically infeasible. If it were practically feasible, the company would extract the entire ore in the period when the tax rate is lowest. The second cost term represents extraction and purification cost, which depends on the total amount of ore extracted in the current period as well as the previous periods, and the metal concentration in it. As the company goes for deep mining, the ore extraction costs may rise. Moreover, the concentration of gold in the ore may vary in different strata. Though it depends on a mine as to how the concentration of metal varies in different strata, we do not make any specific assumptions about it. Rather we combine the costs of the two factors into a single increasing cost function.

\begin{figure}
		\includegraphics[width=\linewidth]{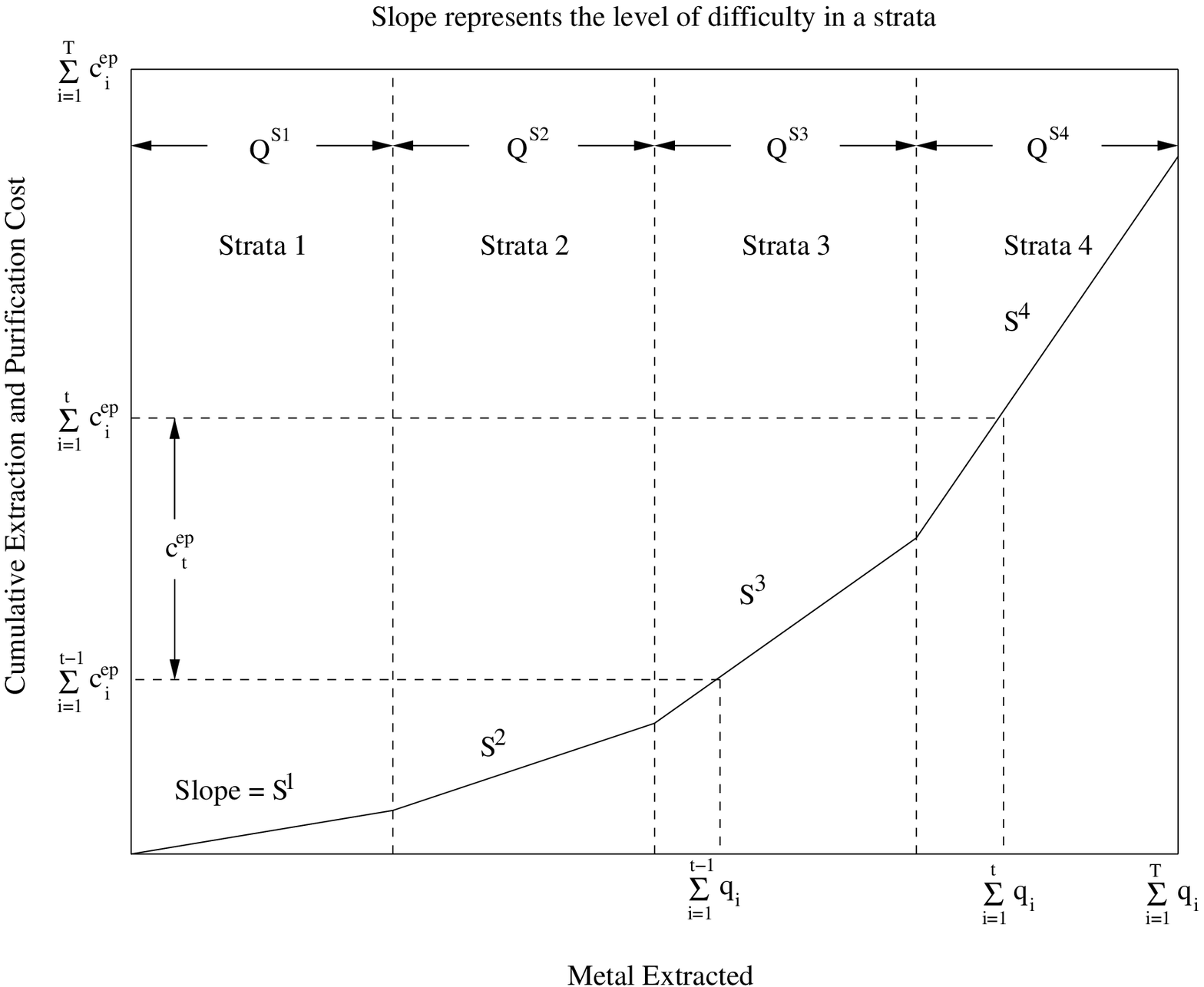}
		\caption[Extraction and Purification Cost]{Cumulative extraction and purification cost against cumulative extraction over periods.}
		\label{fig:cost}
\end{figure}
The extraction rate cost is modeled using a quadratic function as follows:
\begin{equation}
c_{i}^{er}(q_{t},a) = \alpha^{er}(a) q_{t}^{2} + \beta^{er}(a) q_t + \gamma^{er}(a),
\end{equation}
where the coefficients depend upon the choice of technology $a$.
The extraction and purification cost is modeled by a piecewise linear function, as it varies based on the depth of the strata and the purity available in that strata. Instead of providing a mathematical formulation, we provide a graphical representation for this function in Figure \ref{fig:cost}, which is easier to understand. If one decides to mine quantity $q$ of gold from the mine, then the y-axis gives the extraction and purification cost involved in mining that amount of gold from the mine. The figure is partitioned into 4 vertical parts, where each part represents a strata. The amount of gold available in each strata is given by $Q^{S1}$, $Q^{S2}$, $Q^{S3}$ and $Q^{S4}$. A higher slope represents higher costs of extraction. For the function shown in Figure \ref{fig:cost}, the purification and extraction costs increase as one goes from one strata to the other. However, this might not be necessary and entirely depends on the mine. The considered cost function is a piecewise linear curve, which can be completely defined using 8 parameters for four strata; namely $Q^{S1},Q^{S2},Q^{S3},Q^{S4},S^1,S^2,S^3$ and $S^4$.

Next, we provide the parameters for the multi-objective bilevel program, which we attempt to solve using an evolutionary scheme. For the follower, there are four technology alternatives available. Each of the technology alternatives has its own cost and pollution coefficients. The higher the pollution coefficient,  the more is the environmental damage. The mining location has five different strata, and the company operates for 5 periods/years ($T=5$). Table \ref{tab:parameters} contains the parameters corresponding to each technology alternative. The mine has five different strata, and the amount of gold in each strata is given as: $(Q^{S1},Q^{S2},Q^{S3},Q^{S4},Q^{S5})=(20,20,20,20,20)$. The parameters for the price function are: $\alpha = (50,55,60,65,70)$, $\beta = (0.1,0.1,0.1,0.1,0.1)$. The discounting coefficient is assumed to be $r=0.0$ in the computations.

%
%
%

\begin{table}[!h]
\caption{Parameters for each technology alternative.}
\begin{center}
    \begin{tabular}{|l|l|l|l|l|}
        \hline
        Parameters             & $a=1$ & $a=2$ & $a=3$ & $a=4$ \\ \hline
        $k$           & 3     & 5     & 8     & 10     \\ 
        $\alpha^{er}$ & 0.5     & 0.4     & 0.3     & 0.3     \\ 
        $\beta^{er}$  & 5     & 4     & 4     & 2     \\ 
        $\gamma^{er}$  & 10     & 8     & 5     & 5     \\ 
        $S^1$         & 1     & 1     & 0.8     & 0.6     \\ 
        $S^2$         & 1.5     & 1.4     & 1.2     & 0.9     \\ 
        $S^3$         & 2.25     & 1.96     & 1.8     & 1.35     \\ 
        $S^4$         & 3.375     & 2.744     & 2.7     & 2.025     \\
	$S^5$         & 5.063     & 3.842     & 4.05     & 3.038     \\
        \hline
    \end{tabular}
\end{center}
\label{tab:parameters}
\end{table}



\section{Solution Methodology}
\label{sec:algorithm}
\begin{figure}[hbt]
\begin{center}
\epsfig{file=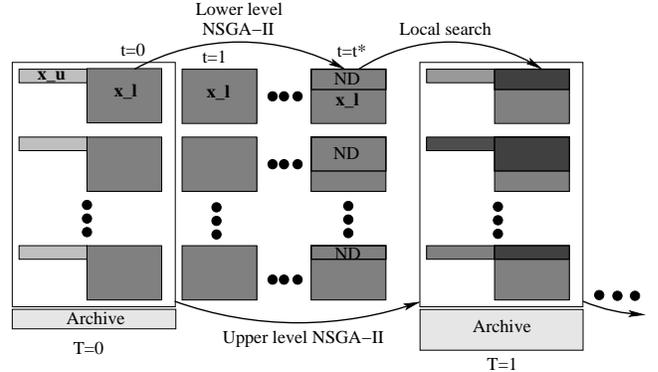,width=1\linewidth}
\end{center}
\caption{A sketch of the H-BLEMO procedure \cite{my-ecj10}.}
\label{fig:algo}
\end{figure}

We solve the extended multi-objective stackelberg model using a recently proposed Hybrid Bilevel Evolutionary Multi-objective Optimization (H-BLEMO) Algorithm \cite{my-ecj10}.
In this section, we briefly outline the working principle of the procedure using a sketch shown in Figure \ref{fig:algo}.

The algorithm starts with an initial population marked with upper level
generation counter $T=0$. It is of size $N_u$ and contains a subpopulation of lower level variable
set $\boldx_l$ for each upper level variable set $\boldx_u$. Initially,
the subpopulation size ($N_l^{(0)}$) is kept identical for each $\boldx_u$
variable set, but it is allowed to change adaptively with generation $T$.
An empty archive $A_0$ is also initialized at the start. 
For each $\boldx_u$, a lower level evolutionary 
optimization is performed on the corresponding subpopulation having variables
$\boldx_l$ alone for a
small number of generations at which the specified 
lower level termination criterion is satisfied.
Thereafter, 
a local search is performed until the local search
termination criterion is
met. The archive is maintained at the upper level, containing
solution vectors $(\boldx_{u_a}, \boldx_{l_a})$, which are optimal at
the lower level and non-dominated at the upper level. 
The solutions in the archive are updated after every lower level call.
The members of the lower level population undergoing a local search
are lower level optimal solutions and hence are assigned an `optimality tag'. These local
searched solutions ($\boldx_l$) are then combined with corresponding $\boldx_u$ variables and become 
eligible to enter the archive if it is
non-dominated when compared to the existing members of the
archive. The dominated members in the archive are then eliminated. The
solutions obtained from the lower level ($\boldx_l$) are combined with corresponding
$\boldx_u$ variables and are processed by the upper level genetic
operators to create a new upper level population. This process is
continued until an upper level termination criterion is satisfied. 
The H-BLEMO algorithm is computationally fast and capable of handling a large number of variables.

\section{Results}
\label{sec:results}
\begin{figure}[!h]
		\includegraphics[width=\linewidth]{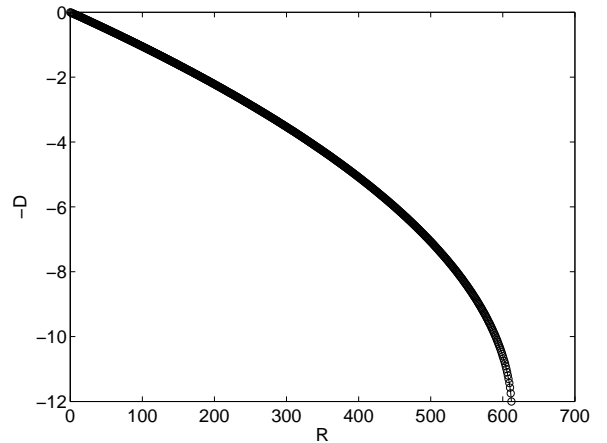}
		\caption[Pareto-optimal Front]{Approximate Pareto-optimal frontier for the government showing the trade-off between tax revenues and environmental pollution obtained using the H-BLEMO approach.}
		\label{fig:gaSimple}
\end{figure}

\begin{figure*}[bht]
\centering
	\begin{minipage}{0.45\textwidth}
		\centering
		\includegraphics[width=\linewidth]{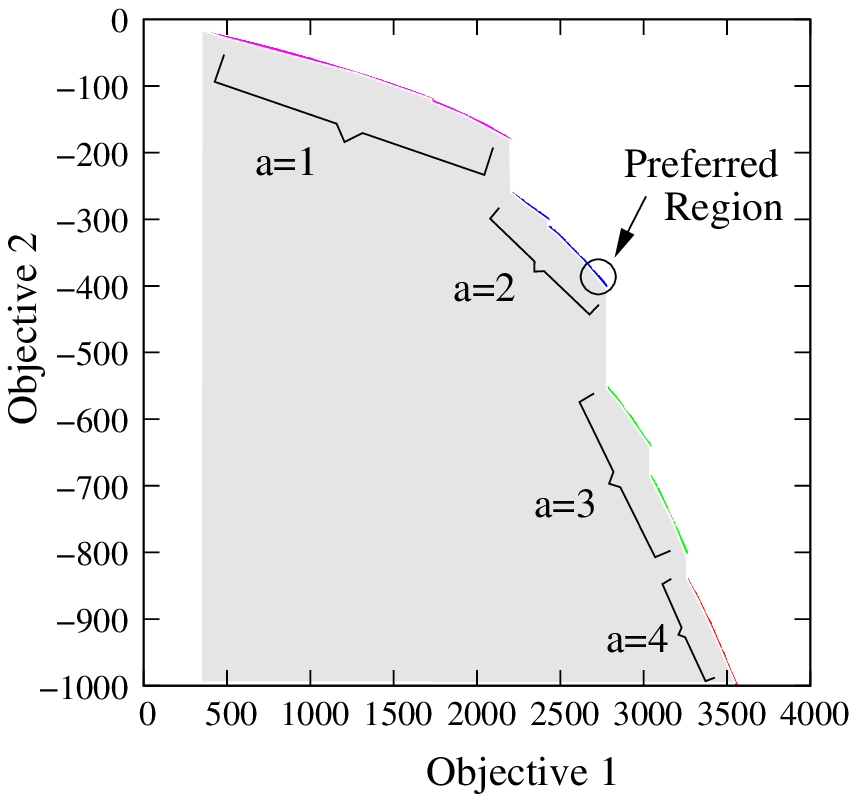}
		\caption[Result 1]{Approximate Pareto-frontier for the extended model taking all technologies into account.}
		\label{fig:result1}
	\end{minipage}\hfill
	\begin{minipage}{0.45\textwidth}
		\centering
		\includegraphics[width=\linewidth]{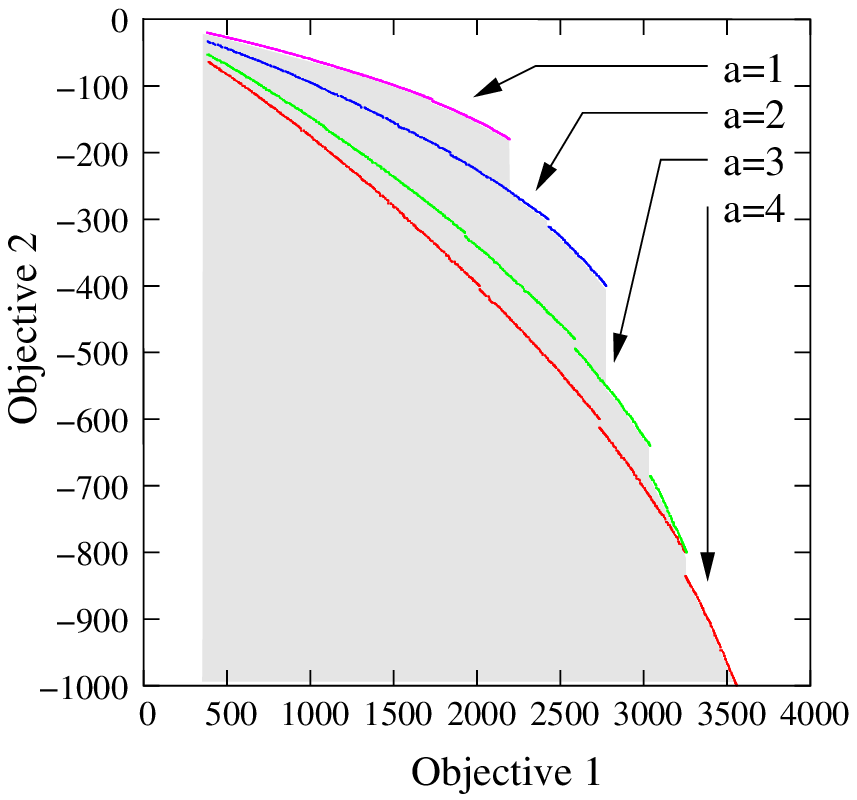}
		\caption[Result 2]{Approximate Pareto-frontiers for the individual technologies considered one at a time.}
		\label{fig:result2}
	\end{minipage}
\end{figure*}

\begin{figure*}[bht]
\centering
	\begin{minipage}{0.49\textwidth}
		\centering
		\includegraphics[width=\linewidth]{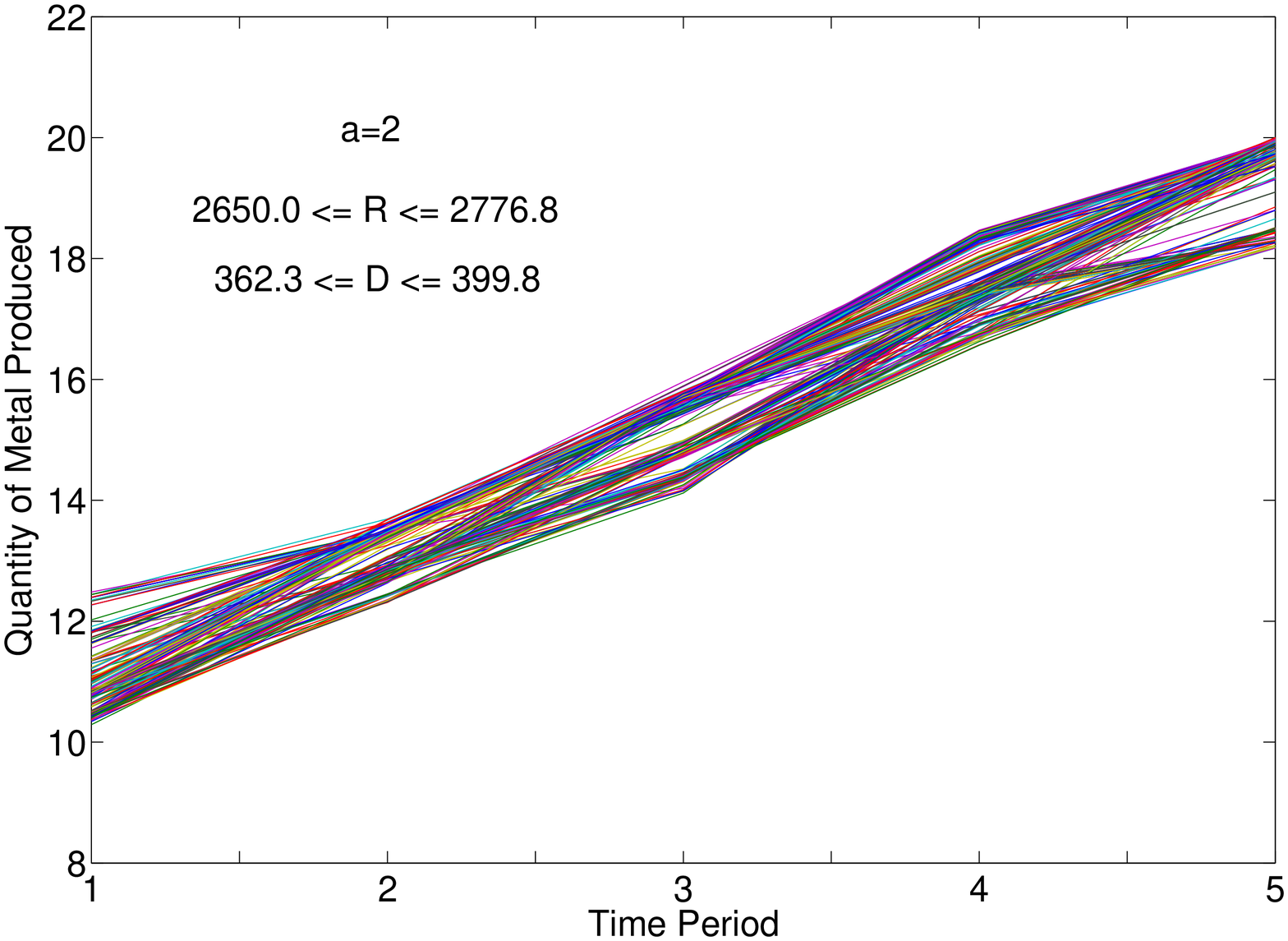}
		\caption[Result 4]{Approximate Pareto-frontier for the extended model taking all technologies into account.}
		\label{fig:result4}
	\end{minipage}\hfill
	\begin{minipage}{0.49\textwidth}
		\centering
		\includegraphics[width=\linewidth]{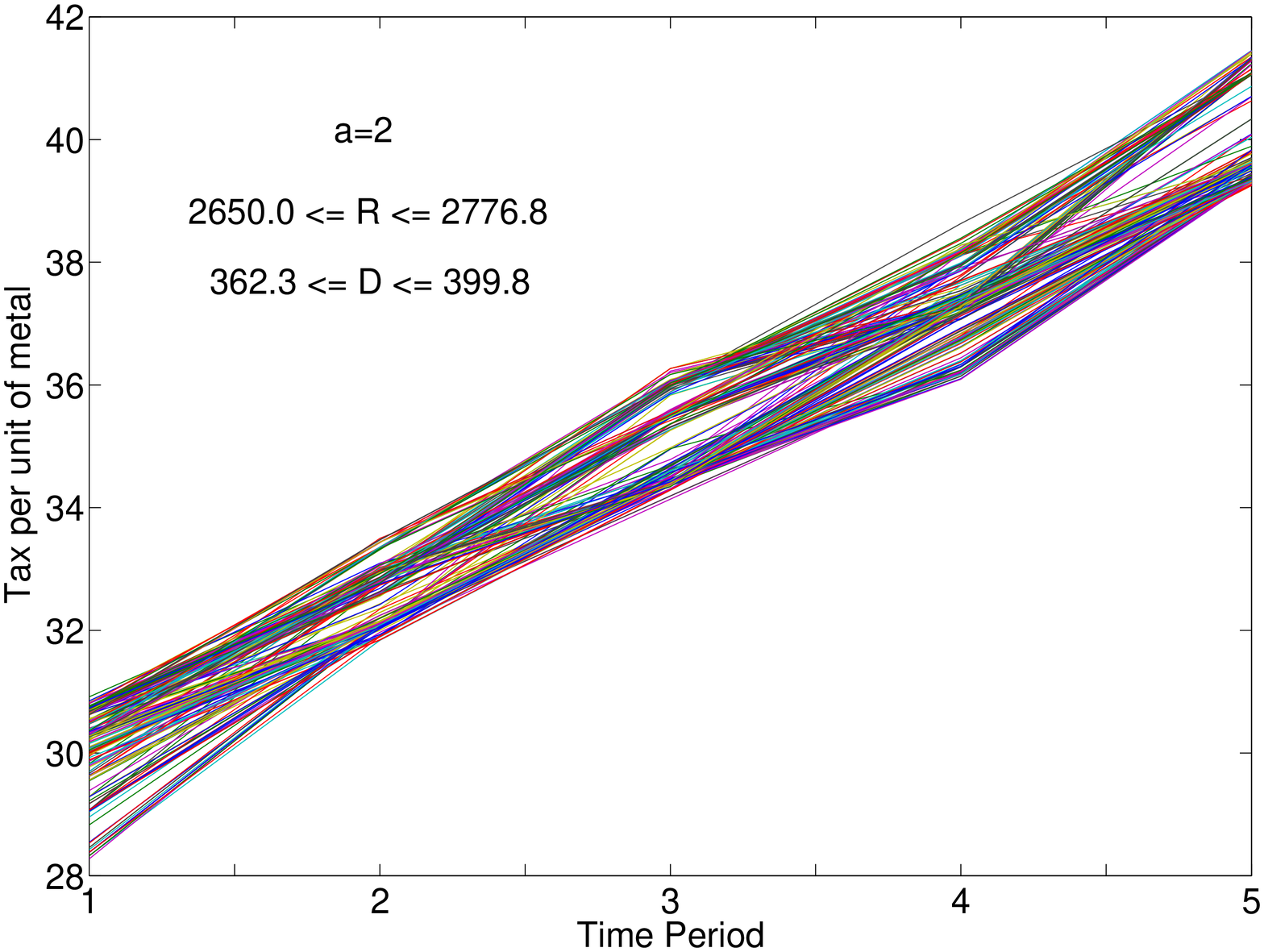}
		\caption[Result 3]{Approximate Pareto-frontiers for the individual technologies considered one at a time.}
		\label{fig:result3}
	\end{minipage}
\end{figure*}

In this section, we present the results obtained using the H-BLEMO algorithm on the simple analytical model as well as the extended multi-objective bilevel model. We execute the H-BLEMO algorithm on the simple analytical model, and we observe that the algorithm is able to converge to the bilevel Pareto-optimal frontier. Figure \ref{fig:gaSimple} shows the approximate Pareto-optimal frontier obtained using the H-BLEMO approach on the simple analytical model. 
Thereafter, we execute the algorithm on the extended bilevel model. Firstly, we consider the extended multi-objective bilevel model as a whole with all of the available technologies and try to generate the approximated Pareto-optimal frontier. The approximate Pareto optimal frontier is shown in Figure \ref{fig:result1}. The parts of the frontier which correspond to different technologies have been marked in the figure. A preferred region is assumed for the government, which is also marked in the figure. We have solved the extended model considering one technology at a time, and generated the Pareto-optimal frontier corresponding to each technology. Different frontiers for each of the technologies are given in Figure \ref{fig:result2}. 
The parts of the different technology frontiers which participate in the combined frontier have been highlighted. The figures represent a maximization frontier, since we handle the models as bilevel multi-objective maximization problems. We are maximizing the revenues and the negative of pollution damage at the upper level, which produces the trade-off frontiers. It can be observed that each of the trade-off frontiers bear a discontinuity, which is caused by the piecewise linear extraction and purification cost. Whenever there is a change in strata, the slope of the piecewise linear extraction cost function changes, which causes discontinuities in the Pareto-frontiers of the respective technologies.
Next, we analyze the solutions in the preferred region. Each solution in the preferred region has its own production level and represents a corresponding tax structure. Figures \ref{fig:result4} and \ref{fig:result3} show the optimal production levels and the optimal taxation strategies during the 5 year time period for the mining company and the government respectively. 

\section{Conclusions and Future Work}
\label{sec:conclusions}
In this paper, we present an application problem with a multi-objective bilevel optimization task. The model involves a Stackelberg competition between a regulating authority and a mining firm. The problem is inspired by a recent controversy in Finland on the harmful impact of gold mining on the environment in the Kuusamo region. We have analyzed the problem from the government's perspective assuming that the government has complete knowledge about the possible actions of the mining company. An analytical example has been provided to give the readers an insight into multi-objective bilevel programming, and then an extended model has been solved using a hybrid bilevel evolutionary multi-objective optimization algorithm.
The solution methodology employed in this problem was effective in handling the bilevel problem and produced a set of trade-off solutions based on which the leader may make a suitable decision.
Our future work on the subject would be focused on incorporating further realism into the model by incorporating a third player into the problem. The player could be one or more tourism companies or the local community, which have been resisting the mining operations in the Kuusamo region. Incorporation of the third entity would allow the government to make an optimal decision, which would be targeted at maximizing the overall welfare. 

\section{Acknowledgments}
The authors wish to thank Liikesivistysrahasto for supporting this study. Authors would also like to thank the Academy of Finland (Grant: 133387) for their support.


\end{document}